\documentstyle[12pt]{article}
\newcommand{\bea}{\begin{eqnarray}}
\newcommand{\eea}{\end{eqnarray}}
\newcommand{\be}{\begin{equation}}
\newcommand{\ee}{\end{equation}}

\begin{document}

\begin{center}
THE RELATIVISTIC PARTICLE WITH CURVATURE AND TORSION OF WORLD TRAJECTORY

\vspace*{1cm}
{\large S.A.Gogilidze,}\\
\it{Tbilisi State University, Tbilisi, University St.9,
380086 Republic of Georgia,} \\
\bigskip
and \\
\bigskip
{\large Yu.S.Surovtsev} \\
\it{Joint Institute for Nuclear Research, Dubna,
Head Post Office.,P.O. Box 79, 101000, Moscow, Russia}
\end{center}
\begin{abstract}
Local symmetries of the action for a relativistic particle with curvature
and torsion of its world curve in the (2+1)-dimensional space-time are
studied. With the help of the method, worked out recently by the authors
(Phys.Rev., D56, 1135, 1142 (1997)), first the local-symmetry transformations are
obtained both in the phase and configuration space. At the classical level,
the dependence of the particle mass on the parameters of curvature and torsion and
the Regge trajectory are obtained. It is shown that the tachyonic sector can
be removed by a proper gauge choice.
\end{abstract}

\section{Introduction}

In recent years, interest in the point relativistic particle has been
revived by investigation of the action depending on the curvature and/or
torsion of the world trajectory of that particle (see, {\it e.g.}
\cite{Kuz.Plyu.} and refs. therein). This relates, on the one hand, to the
development of the string theory: the above-mentioned object can be considered
as a one-dimensional analog of the relativistic string with rigidity
\cite{Polyakov}. On the other hand, such models seem to continue realizing an
old idea of modelling the particle spin without introducing additional spin
variables \cite{Casimir}. Note also a connection of the indicated models with
the problem of describing planar phenomena: the high-temperature
superconductivity \cite{Polyakov2} and the fractional quantum Hall effect
\cite{Gerbert}.

The present work is devoted to an investigation of the action of a massive
relativistic particle with curvature and torsion in (2+1) dimensions. Here
there are the first- and second-class constraints, both primary and secondary
(see, {\it e.g.} \cite{Kuz.Plyu.,Nest.1}). Therefore, already the initial
stage of constructing the theory needs a complete separation of constraints of
the first and second class and finding gauge transformations. This can be made
only with the help of the methods, recently proposed in papers
\cite{cgs-pr1}-\cite{cgs-IJMP}. An explicit form of gauge transformations is
required, {\it e.g.}, in eliminating unphysical degrees of freedom ({\it i.e.},
a reduction of the theory), carried out before or after the quantization (the
schemes of covariant and canonical quantization).

The layout of the paper is as follows. In Sec. 2, the system under
investigation is considered in the framework of the generalized Hamiltonian
formalism by Dirac for theories with higher derivatives \cite{Ostr.}, and by
the method, suggested in Ref.\cite{cgs-pr1}, we separate completely the
constraints of the first and second class and obtain a canonical set of
constraints. In Sec. 3, we obtain gauge transformations for the considered
system both in the phase and configuration space and the Noether identity.
In Sec. 4, the dependence of the particle mass on the parameters of
curvature and torsion and the Regge trajectory are obtained. It is shown that
the tachyonic sector can be removed by a proper gauge choice.\\

\section{Generalized Hamiltonian formalism}

So, let us consider a relativistic particle with curvature and
torsion of its world curve in the (2+1)-dimensional space-time. Given a
parametrization of the world trajectory $x^\mu(\tau)~(\mu=0,1,2)$, the action
can be written in the form
\begin{equation} \label{S}
S=-m\int d\tau\sqrt{{\dot x}^2} - \alpha\int d\tau\frac{\sqrt{({\dot x}
{\ddot x})^2-{\dot x}^2{\ddot x}^2}}{{\dot x}^2}-\beta\int d\tau
\sqrt{{\dot x}^2}\frac{\varepsilon_{\mu\nu\sigma}\stackrel{.}{x}^\mu
\stackrel{..}{x}^\nu \stackrel{\ldots}{x}^{\sigma}}{({\dot x}{\ddot x})^2-
{\dot x}^2{\ddot x}^2},
\end{equation}
where ~${\dot x}\equiv dx(\tau)/d\tau,~{\dot x}^2>0,$ dimensionless
parameters $\alpha$ and $\beta$ characterize the contributions of curvature
($k$) and torsion ($\kappa$), respectively, $m$ is a parameter of the mass
dimension, $\varepsilon_{\mu\nu\rho}$ is a completely antisymmetric unit
tensor ($\varepsilon_{012}=1$), the Lorentz metric with signature $(+,-,-)$ is
used. In the following we shall use the notation
\begin{equation} \label{eps-nota}
\varepsilon_{\mu\nu\rho}{a}^\mu{b}^\nu{c}^{\rho}=\varepsilon(abc),~~~~~
\varepsilon_{\mu\nu\rho}{a}^\nu{b}^{\rho}=\varepsilon_{\mu}(ab).
\end{equation}
We see that the action \ref{S} depends on the derivatives of the particle
coordinates with respect to $\tau$ up to and including the third order.
Therefore, when passing to the generalized Hamiltonian formalism for theories
with higher derivatives, the canonical variables are to be introduced
according to the Ostrogradsky method \cite{Ostr.}
\begin{eqnarray} \label{can.variables}
&&q_1=x,~~~~~q_2=\dot x,~~~~~q_3=\ddot x,\\
&&p_1=-\frac{\partial L}{\partial{\dot x}}-{\dot p}_2,~~~~p_2=-
\frac{\partial L}{\partial{\ddot x}}-{\dot p}_3,~~~~
p_3=-\frac{\partial L}{\partial{\stackrel{\ldots}{x}}},
\end{eqnarray}
where $L$ is the Lagrangian in \ref{S}. The canonical Hamiltonian has the form
\begin{equation} \label{H_c}
H_c=-p_1{\dot x}-p_2{\ddot x}-p_3{\stackrel{\ldots}{x}}-L=-p_1q_2-p_2q_3+
m\sqrt{q_2^2}+\alpha\frac{\sqrt{g}}{{q_2}^2}.
\end{equation}
The Poisson brackets (PB) are defined in the following way
\begin{equation} \label{PB}
\{A,B\}=\sum_{a=1}^3\left(\frac{\partial A}{\partial {p_a^\mu}}\frac{\partial B}
{\partial {q_{a\mu}}}-\frac{\partial A}{\partial {q_a^\mu}}\frac{\partial B}
{\partial {p_{a\mu}}}\right).
\end{equation}
From the definition of the momentum $p_3$ three primary constraints are
derived
\begin{equation} \label{p_3}
p_{3\mu}-\beta\frac{\sqrt{q_2^2}}{g}\varepsilon_\mu(q_2q_3)=0,~~~~\mu=0,1,2,
\end{equation}
where $~g=(q_2q_3)^2-q_2^2 q_3^2$. Making use of the orthonormal basis in a
three-dimensional space-time
\begin{equation} \label{basis}
n_1^\mu=\frac{q_2^\mu}{\sqrt{q_2^2}},~~~~
n_2^\mu=\sqrt{\frac{q_2^2}{g}}\left(q_3^\mu-\frac{q_2q_3}{q_2^2}
q_2^\mu\right),~~~~
n_3^\mu=\frac{\varepsilon^\mu(q_2q_3)}{\sqrt{g}},
\end{equation}
and the completeness condition $~n_1^\mu n_1^\nu -n_2^\mu n_2^\nu -
n_3^\mu n_3^\nu=g^{\mu\nu}$, it is convenient to decompose the vector \ref{p_3}
in this basis and to pass from three primary constraints \ref{p_3} to the
equivalent constraints (the projections of \ref{p_3} onto the vectors
\ref{basis})
\begin{equation} \label{prim.constr.}
\phi_1^1=p_3 q_2,~~~~\phi_2^1=p_3 q_3,~~~~ \phi_3^1=\varepsilon(p_3q_2q_3)+
\beta\sqrt{q_2^2}.
\end{equation}
Determinant of the matrix of the transformation to the constraint set
\ref{prim.constr.} equals $~-g$.  When investigating self-consistency of the
theory, we apply the Dirac scheme for breeding constraints. From the
conditions of the time conservation of constraints
${\dot \phi}_\alpha^1=\{\phi_\alpha^1,H_c+u_\alpha\phi_\alpha^1\}~
\stackrel{\Sigma_1}{=}0~$ ($\alpha=1,2,3; ~u_\alpha$ are the Lagrange
multipliers; ~$\stackrel{\Sigma_1}{=}$ means this equality to hold on the
surface of the primary constraints $\Sigma_1$) we obtain the secondary
constraints
\begin{equation} \label{sec.constr.}
\phi_1^2=p_2 q_2,~~~~\phi_2^2=p_2 q_3-\alpha\frac{\sqrt{g}}{{q_2}^2},~~~~
\phi_3^2=\varepsilon(p_2q_2q_3)-\beta\frac{q_2q_3}{\sqrt{q_2^2}}.
\end{equation}
From the conditions ${\dot \phi}_\alpha^2~\stackrel{\Sigma_2}{=}0~
(\alpha=1,2)$ two ternary constraints are obtained
\begin{equation} \label{tert.constr.}
\phi_1^3=p_1 q_2-m\sqrt{q_2^2},~~~~~\phi_2^3=p_1 q_3-
m\frac{q_2q_3}{\sqrt{q_2^2}},
\end{equation}
and the equation  ${\dot \phi}_3^2~\stackrel{\Sigma_2}{=}0$  serves for
determining the Lagrangian multiplier $u_3$
\begin{equation} \label{u_3}
u_3=\frac{1}{\alpha\sqrt{g}}\left[\varepsilon(p_1q_2q_3)-
\frac{\beta g}{(q_2^2)^{3/2}}\right].
\end{equation}
Further we obtain
$${\dot\phi}_1^3~=~\phi_2^3$$
and from ~${\dot \phi}_2^3~\stackrel{\Sigma_3}{=}0$~ the quaternary constraint
\begin{equation} \label{quater.constr.}
\phi_2^4=\frac{\sqrt{g}}{\alpha}\left[p_1^2-m^2+
\sqrt{g}~\frac{\alpha m+\beta\sigma\sqrt{m^2-p_1^2}}{(q_2^2)^{3/2}}\right],~~~~~~
\sigma=\mbox{sign}~\varepsilon(p_1q_2q_3),
\end{equation}
from the conservation (in time) conditions of which the Lagrangian multiplier
$u_2$
\begin{equation} \label{u_2}
u_2=-3\frac{q_2q_3}{q_2^2}
\end{equation}
is determined. In eq.\ref{quater.constr.} it is taken into account that
$$[\varepsilon(p_1q_2q_3)]^2~\stackrel{\Sigma_3}{=}~g(m^2-p_1^2)$$.

So, the complete set of constraints is obtained with the following PB among
them:
\begin{eqnarray} \label{PB-original}
&&\{\phi_1^1,\phi_2^1\}=\phi_1^1,~~~~~\{\phi_1^1,\phi_3^1\}=0,~~~~~
\{\phi_2^1,\phi_3^1\}=0,\nonumber\\
&&\{\phi_1^2,\phi_1^1\}=\phi_1^1,~~~~~\{\phi_1^2,\phi_2^1\}=0,~~~~~
\{\phi_1^2,\phi_3^1\}=\phi_3^1,\nonumber\\
&&\{\phi_1^2,\phi_2^2\}=-\phi_2^2,~~~\{\phi_1^2,\phi_3^2\}=0,\nonumber\\
&&\{\phi_1^3,\phi_1^1\}=0,~~~~~\{\phi_1^3,\phi_2^1\}=0,~~~~~
\{\phi_1^3,\phi_3^1\}=0,~~~~~\{\phi_1^3,\phi_1^2\}=-\phi_1^3,\nonumber\\
&&\{\phi_1^3,\phi_2^2\}=-\phi_2^3,~~~~~
\{\phi_1^3,\phi_3^2\}=-\varepsilon(p_1q_2q_3),~~~~~
\{\phi_1^3,\phi_2^3\}=0,\nonumber\\
&&\{\phi_2^2,\phi_1^1\}=\phi_2^1-\phi_1^2,~~~
\{\phi_2^2,\phi_2^1\}=-\phi_2^2,~~~\{\phi_2^2,\phi_3^1\}=-\phi_3^2,
\nonumber\\
&&\{\phi_2^3,\phi_1^1\}=-\phi_1^3,~~~~\{\phi_2^3,\phi_2^1\}=-\phi_2^3,~~~~
\{\phi_2^3,\phi_3^1\}=-\varepsilon(p_1q_2q_3),\\
&&\{\phi_2^3,\phi_1^2\}=0,~~~~~~\{\phi_2^3,\phi_2^2\}=
-\frac{mg}{(q_2^2)^{3/2}},~~~~~~\{\phi_2^3,\phi_3^2\}=0,\nonumber\\
&&\{\phi_2^4,\phi_1^1\}=0,~~~~\{\phi_2^4,\phi_2^1\}=-\phi_2^4+
\frac{g}{(q_2^2)^{3/2}}\left(m+
\frac{\beta}{\alpha}\sigma\sqrt{m^2-p_1^2}\right),\nonumber\\
&&\{\phi_2^4,\phi_3^1\}=0,~~~~
\{\phi_2^4,\phi_1^2\}=-\phi_2^4+2\frac{g}{(q_2^2)^{3/2}}\left(m+
\frac{\beta}{\alpha}\sigma\sqrt{m^2-p_1^2}\right),\nonumber\\
&&\{\phi_2^4,\phi_2^2\}=3~\frac{q_2q_3}{q_2^2}\frac{g}{(q_2^2)^{3/2}}\left(m+
\frac{\beta}{\alpha}\sigma\sqrt{m^2-p_1^2}\right),\nonumber\\
&&\{\phi_2^4,\phi_3^2\}=0,~~~~~~\{\phi_2^4,\phi_1^3\}=0,~~~~~~
\{\phi_2^4,\phi_2^3\}=0,\nonumber\\
&&\{\phi_3^2,\phi_1^1\}=\phi_3^1,~~~~\{\phi_3^2,\phi_2^1\}=-\phi_3^2,\nonumber\\
&&\{\phi_3^2,\phi_3^1\}=\alpha\sqrt{g}-q_3^2\phi_1^1+q_2^2\phi_2^2+
q_2q_3(\phi_2^1-\phi_1^2),~~~~
\{\phi_2^2,\phi_3^2\}=\frac{\beta g}{(q_2^2)^{3/2}}.\nonumber
\end{eqnarray}
We see that from three chains of the constraints the two ones are candidates
for second-class constraints. Making use of the method developed in
Ref.\cite{cgs-pr1}, let us separate the constraints into the first- and
second-class ones and pass to an equivalent canonical set of constraints.
For this purpose, we perform the equivalence transformation to obtain the
desired structure of the second-class constraints of the canonical set and
then completely to separate the first-class constraints:
\begin{equation} \label{Psi_2,3}
\Psi_2^2=\phi_2^2+a_1\phi_2^1+a_2\phi_3^1,~~~~~\Psi_2^3=\phi_2^3+a_3\phi_3^2,
\end{equation}
where the coefficients $a_1,a_2$ and $a_3$ are determined to fulfil the
requirements
\begin{equation} \label{to-Psi}
\{\Psi_2^2,\Psi_3^2\}\stackrel{\Sigma}{=} 0,~~~~
\{\Psi_2^2,\Psi_2^4\}\stackrel{\Sigma}{=} 0,~~~~
\{\Psi_2^3,\Psi_3^1\}\stackrel{\Sigma}{=} 0
\end{equation}
($\stackrel{\Sigma}{=}$ means this equality to hold on the surface of
all constraints $\Sigma$). Now we carry out the following equivalence
transformation
\begin{equation}
\Phi_1^2=\phi_1^2+b_1\phi_2^1,~~~~~
\Phi_1^3=\phi_1^3+b_2\phi_3^1+b_3\Psi_2^2
\end{equation}
with coefficients $b_1,b_2$ and $b_3$, determined to fulfil the requirements
\begin{equation} \label{to-Phi_1}
\{\Phi_1^2,\Psi_2^4\}\stackrel{\Sigma}{=} 0,~~~~
\{\Phi_1^3,\Psi_3^2\}\stackrel{\Sigma}{=} 0,~~~~
\{\Phi_1^3,\Psi_2^3\}\stackrel{\Sigma}{=} 0.
\end{equation}
The obtained set of constraints
\begin{eqnarray}
&&\Phi_1^1=\phi_1^1,~~\Phi_1^2=\phi_1^2+2\phi_2^1,\nonumber\\
&&\Phi_1^3=\phi_1^3+3\frac{q_2q_3}{q_2^2}\phi_2^1+
\left[\frac{\beta}{\alpha}\frac{\sqrt{g}}{(q_2^2)^{3/2}}-
\frac{\varepsilon(p_1q_2q_3)}{\alpha\sqrt{g}}\right]\phi_3^1+\phi_2^2,~~~~~~~
\label{Phi_1}\\
&&\Psi_2^1=\phi_2^1,~~\Psi_2^2=\phi_2^2+3\frac{q_2q_3}{q_2^2}\phi_2^1+
\frac{\beta}{\alpha}\frac{\sqrt{g}}{(q_2^2)^{3/2}}\phi_3^1,\nonumber\\
&&\Psi_2^3=\phi_2^3+\frac{\varepsilon(p_1q_2q_3)}{\alpha\sqrt{g}}\phi_3^2,~~
\Psi_2^4=\phi_2^4,~~~~~~~\label{Psi_2}\\
&&\Psi_3^1=\phi_3^1,~~\Psi_3^2=\phi_3^2 \label{Psi_3}
\end{eqnarray}
is canonical, because the only non-vanishing (on $\Sigma$) PB among them are
\begin{equation} \label{non-0-PB}
\{\Psi_2^1,\Psi_2^4\}\stackrel{\Sigma}{=}\{\Psi_2^2,\Psi_2^3\}
\stackrel{\Sigma}{=}\frac{g}{(q_2^2)^{3/2}}\left(m+
\frac{\beta}{\alpha}\sigma\sqrt{m^2-p_1^2}\right),~~~~
\{\Psi_3^1,\Psi_3^2\}=-\alpha\sqrt{g},
\end{equation}
{\it i.e.}, on $\Sigma$ the maximal partition of the set of constraints is
achieved:  each second-class constraint of the final set has the vanishing
(on the constraint surface) PB with all the constraints of the system except
one, and the first-class constraints have the vanishing PB with all the
constraints. Therefore, there are three constraints of the first class
\ref{Phi_1} and two chains of the second-class constraints \ref{Psi_2},
\ref{Psi_3} (second-class constraints are denoted by the letter $\Psi$).

The total Hamiltonian has now the form
\begin{equation} \label{H_T}
H_T=-\Phi_1^3+u_1\Phi_1^1,
\end{equation}
{\it i.e.}, is a function of the first-class.
Note that the secondary constraints of the first class have the following PB
with the primary constraint $\Phi_1^1$
\begin{equation} \label{non-ideal}
\{\Phi_1^2,\Phi_1^1\}=-\Phi_1^1,~~~~~~
\{\Phi_1^3,\Phi_1^1\}=-3\frac{q_2q_3}{q_2^2}\Phi_1^1-\Phi_1^2.
\end{equation}
The second PB \ref{non-ideal} are not of the form needed for obtaining
local-symmetry transformations -- the right-hand side must be expressed only
through $\Phi_1^1$ (first-class primary constraints are to be the ideal of
quasi-algebra of all the first-class constraints \cite{cgs-pr2}). Therefore,
instead of $\Phi_1^3$ in \ref{Phi_1}, the equivalent constraint will be used
\begin{equation} \label{Phi_1^3}
\overline{\Phi}_1^3=\Phi_1^3-\frac{q_2q_3}{q_2^2}~\Phi_1^2,
\end{equation}
for which
\begin{equation} \label{ideal}
\{\overline{\Phi}_1^3,\Phi_1^1\}=-2\frac{q_2q_3}{q_2^2}\Phi_1^1.
\end{equation} \\

\section{Local-symmetry transformations}

Now we can say that, according to the theorem proved in
Ref.\cite{cgs-pr2}, the action \ref{S} is quasi-invariant under the
one-parameter quasigroup of local-symmetry transformations. We shall derive
these transformations by the method developed in Ref.\cite{cgs-pr2,cgs-IJMP},
where it has been proved that for the obtained canonical set of constraints
the generator of local-symmetry transformations is to be the form
\begin{equation} \label{G}
G=\varepsilon_1^1~\Phi_1^1+\varepsilon_1^2~\Phi_1^2+
\varepsilon_1^3~\overline{\Phi}_1^3,
\end{equation}
where the coefficients $\varepsilon_1^{m_1}~ (m_1=1,2,3)$ are the solution of
the system of equations
\begin{equation} \label{eq:eps}
\dot \varepsilon_1^{m_1}+
\varepsilon_1^{m_1^\prime}g_{1~~1}^{m_1^\prime m_1}=0,~~~~~~
m_1^\prime =m_1-1,\cdots,3
\end{equation}
with the functions $g_{1~~1}^{m_1^\prime m_1}$ determined from PB
\begin{equation} \label{PB-Phi-H}
\{\Phi_1^{m_1},H\} =g_{1~~1}^{m_1 m_1^\prime}~\Phi_1^{m_1^\prime},~~~~~
m_1^\prime=1,\cdots,m_1+1
\end{equation}
(here $H=H_c+u_2\Psi_2^1+u_3\Psi_3^1$ with $u_2$ and $u_3$ given by eqs.
\ref{u_2} and \ref{u_3}).
As ~$g_{1~1}^{3~3}=q_2q_3/q_2^2,~g_{1~1}^{2~3}=-1,~g_{1~1}^{3~2}=-q_3^2/q_2^2,
~g_{1~1}^{2~2}=-q_2q_3/q_2^2,~g_{1~1}^{1~2}=-1$, ~the system of equations
\ref{eq:eps} for determining $\varepsilon_1^{m_1}$ becomes
\begin{eqnarray} \label{eps:eq.}
&&\dot{\varepsilon}_1^3+\frac{q_2q_3}{q_2^2}\varepsilon_1^3-
\varepsilon_1^2=0,\nonumber\\
&&\dot{\varepsilon}_1^2-\frac{q_3^2}{q_2^2}\varepsilon_1^3-
\frac{q_2q_3}{q_2^2}\varepsilon_1^2-\varepsilon_1^1=0.
\end{eqnarray}
Denoting ~$\varepsilon_1^3\equiv \lambda$,~ we obtain
\begin{equation} \label{eps:solut.}
\lambda_1^2=\dot{\lambda}+\lambda\frac{q_2q_3}{q_2^2},~~~~
\lambda_1^1=\ddot{\lambda}+\frac{\lambda}{q_2^2}\left[q_2{\dot{q}}_3-
\frac{q_2q_3}{q_2^2}(2q_2{\dot{q}}_2+q_2q_3)\right].
\end{equation}
We see that the quantity $G$ in \ref{G} depends on ${\dot{q}}_3$; therefore,
for the desired local-symmetry transformations being canonical it is necessary
to extend the phase space in the following way: \\
Define the coordinates ~$Q_i=q_i~~(i=1,2,3),~~Q_4={\dot{q}}_3$; then their
conjugate momenta, calculated in accordance with the formula of theories with
higher derivatives \cite{Ostr.}, are
~$P_i=p_i~~(i=1,2,3),~~P_4=0$. The generalized momentum $P_4$ is an extra
primary constraint of the first class.
In the extended phase space the total Hamiltonian is written down as
\begin{equation} \label{enlarged-H}
\overline{H}_T = H_T(Q_i,P_i) + {\bar u}P_4 ,
\end{equation}
where $H_T$ is of the same form as in the initial phase space \ref{H_T} and
${\bar u}$ is an arbitrary function of the time.
From \ref{enlarged-H} we may conclude that there do not appear additional
secondary constraints corresponding to $P_4$. The set of constraints in the
extended phase space remains the same as in the initial phase space, obeys
the same algebra and does not depend on new coordinates and momenta as $H_T$
does.

We shall seek a generator $\overline{G}$ in the extended phase space in a
form analogous to that in the initial phase space. Then from the requirements
of quasi-invariance of the action
\begin{equation} \label{enlarged-S}
\overline{S}=\int_{\tau_1}^{\tau_2}d\tau\left[P_1Q_2+P_2Q_3+P_3Q_4+
P_4{\dot Q_4} - \overline{H}_T \right]
\end{equation}
under the transformations generated by $\overline{G}$, we obtain the same
relations \ref{eps:solut.} for determining $\varepsilon_1^{m_1}$.
As a result, we find the expression for $\overline{G}$:
\begin{eqnarray} \label{enl.G}
\overline{G} &=& \ddot{\lambda}(P_3Q_2)+2\dot{\lambda}(P_2Q_2+2P_3Q_3)+
\lambda\biggl\{P_1Q_2+P_2Q_3+3P_3Q_3\frac{Q_2Q_3}{Q_2^2}+\nonumber\\
&&\frac{P_3Q_2}{Q_2^2}\left[Q_2Q_4-3\frac{(Q_2Q_3)^2}{Q_2^2}\right]-
m\sqrt{Q_2^2}-\alpha\frac{\sqrt{g}}{Q_2^2}+\\
&&\left[\frac{\beta}{\alpha}\frac{\sqrt{g}}{(Q_2^2)^{3/2}}-
\frac{\varepsilon(P_1Q_2Q_3)}{\alpha\sqrt{g}}\right]
\left[\varepsilon(P_3Q_2Q_3)+\beta\sqrt{Q_2^2}\right]\biggr\}+
\overline{\lambda} P_4.\nonumber
\end{eqnarray}
Note that the obtained generator \ref{enl.G} satisfies the group property
\begin{equation} \label{enl.G:group}
\{\overline{G}_1,\overline{G}_2\}=\overline{G}_3,
\end{equation}
where the transformation $\overline{G}_3$ \ref{enl.G} is realized by carrying
out two successive transformations $\overline{G}_1$ and $\overline{G}_2$
\ref{enl.G}. Now the local-symmetry transformations of the coordinates in the
extended one are of the form
\begin{eqnarray} \label{enl.deltaQP}
&&\delta Q_1^\mu=\{Q_1^\mu,\overline{G}\}=-\lambda\left\{Q_2^\mu-
\frac{\varepsilon^\mu(Q_2Q_3)}{\alpha\sqrt{g}}\left[\varepsilon(P_3Q_2Q_3)+
\beta\sqrt{Q_2^2}\right]\right\},\nonumber\\
&&\delta Q_2^\mu=-\dot{\lambda}Q_2^\mu- \lambda Q_3^\mu, \nonumber\\
&&\delta Q_3^\mu=-\ddot{\lambda}Q_2^\mu-2\dot{\lambda}Q_3^\mu-
\lambda\biggl\{\frac{Q_2^\mu}{Q_2^2}\left[Q_2Q_4-3\frac{(Q_2Q_3)^2}{Q_2^2}
\right]+\nonumber\\
&&~~~~~~~~~3Q_3^\mu\frac{Q_2Q_3}{Q_2^2}\varepsilon_\mu (Q_2Q_3)
\left[\frac{\beta}{\alpha}\frac{\sqrt{g}}{(Q_2^2)^{3/2}}-
\frac{\varepsilon(P_1Q_2Q_3)}{\alpha\sqrt{g}}\right]\biggr\},\nonumber\\
&&\delta Q_4^\mu=-\overline{\lambda}^\mu,\\
&&\delta P_1^\mu=\{P_1^\mu,\overline{G}\}=0,\nonumber\\
&&\delta P_2^\mu=f(\ddot{\lambda},\dot{\lambda},\lambda;Q_2,Q_3,Q_4;
P_1,P_2,P_3),~~~\delta P_3^\mu=h(\dot{\lambda},\lambda;Q_2,Q_3;
P_1,P_2,P_3),\nonumber\\
&&\delta P_4^\mu=\lambda\frac{P_3Q_2}{Q_2^2}Q_2^\mu,\nonumber
\end{eqnarray}
here the functions $f$ and $h$ are rather cumbersome. One can verify that to
within quadratic terms in $\delta Q_i$ and $\delta P_j$
\begin{equation} \label{PB:QP}
\{Q_i+\delta Q_i ,P_j+\delta P_j\}=\delta_{ij},
\end{equation}
{\it i.e.}, the obtained infinitesimal gauge transformations are canonical in
the extended (by Ostrogradsky) phase space.

We see from \ref{enl.deltaQP} that the vector $P_1^\mu$ is gauge-invariant, and
it is conserved also in the time evolution.

The corresponding transformations of local symmetry in the Lagrangian
formalism are determined in the following way:
\begin{equation} \label{x-dot-x}
\delta x(\tau)=\{q_1(\tau),G\}\left(q_1=x,q_2=\dot{x},q_3=\ddot{x},
p_3=-\frac{\partial L}{\partial{\stackrel{\ldots}{x}}}\right),~~~
\delta{\dot x}(\tau)=\frac{d}{d\tau}\delta x(\tau).
\end{equation}
We obtain
\begin{equation} \label{delta-x}
\delta x^\mu=-\lambda~{\dot{x}}^\mu.
\end{equation}
One can see that the Lagrangian \ref{S} is quasi-invariant under these
transformations. The corresponding Euler - Lagrange equation is
\begin{equation} \label{Euler-Lagr.eq.}
\frac{d}{d\tau}\frac{\partial L}{\partial{\dot x}}-
\frac{d^2}{d\tau^2}\frac{\partial L}{\partial{\ddot x}}+
\frac{d^3}{d\tau^3}\frac{\partial L}{\partial{\stackrel{\ldots}{x}}}=0.
\end{equation}
Under the transformations \ref{x-dot-x}, \ref{delta-x}
\begin{eqnarray} \label{delta-L}
\delta L &=& \frac{\partial L}{\partial{\dot x}_\mu}\delta{\dot x}_\mu+
\frac{\partial L}{\partial{\ddot x}_\mu}\delta{\ddot x}_\mu+
\frac{\partial L}{\partial{\stackrel{\ldots}{x}}_\mu}
\delta{\stackrel{\ldots}{x}}_\mu=
\frac{\delta L}{\delta x_\mu}{\delta x_\mu}+  \nonumber\\
&&\frac{d}{d\tau}\biggl[\left(\frac{\partial L}{\partial{\dot x}_\mu}-
\frac{d}{d\tau}\frac{\partial L}{\partial{\ddot x}_\mu}+
\frac{d^2}{d\tau^2}\frac{\partial L}{\partial{\stackrel{\ldots}{x}}_\mu}
\right){\delta x_\mu}+\nonumber\\
&&\left(\frac{\partial L}{\partial{\ddot x}_\mu}-
\frac{d}{d\tau}\frac{\partial L}{\partial{\stackrel{\ldots}{x}}_\mu}\right)
\delta{\dot x}_\mu+\frac{\partial L}{\partial{\stackrel{\ldots}{x}}_\mu}
{\delta\ddot x}_\mu\biggr].
\end{eqnarray}
We obtain the Noether identity
\begin{equation} \label{Noether}
\left(\frac{d}{d\tau}\frac{\partial L}{\partial{\dot x}_\mu}-
\frac{d^2}{d\tau^2}\frac{\partial L}{\partial{\ddot x}_\mu}+
\frac{d^3}{d\tau^3}\frac{\partial L}{\partial{\stackrel{\ldots}{x}}_\mu}
\right){\dot x}_\mu=0.
\end{equation}
Then we have ~$\delta L=-(d/d\tau)(\lambda L)$. \\
In particular, curvature ($k$) and torsion ($\kappa$) are transformed under
the transformations \ref{x-dot-x} and \ref{delta-x} as
\begin{equation} \label{transf:curv-tors.}
\delta k=-\lambda~\dot{k},~~~~~~~\delta\kappa=-\lambda~\dot{\kappa};
\end{equation}
therefore, their contributions to the action $S$ \ref{S} are quasi-invariant
under \ref{x-dot-x}:
\begin{equation} \label{contrib:curv-tors.}
\delta L[k]=\alpha\frac{d}{d\tau}(\lambda\sqrt{{\dot x}^2}k),~~~~~~~
\delta L[\kappa]=\beta\frac{d}{d\tau}(\lambda\sqrt{{\dot x}^2}\kappa).
\end{equation}\\

\section{Spin-mass-curvature relation}

Consider some physical conclusions (first on the classical level)
from the obtained set of constraints and the transformations of local
symmetry. We have already seen that the integral of motion $P_1$ is
gauge-invariant. From invariance of the action \ref{enlarged-S} under
translations we obtain that $P_1$ is the energy-momentum vector. From
invariance of the action \ref{enlarged-S} under rotations we obtain that the
conserved angular momentum tensor is
\begin{equation} \label{ang.mom.tensor}
M_{\mu\nu}=\sum_{a=1}^{4}(Q_{a\mu}P_{a\nu}-P_{a\mu}Q_{a\nu}).
\end{equation}
This quantity is gauge-invariant in the initial phase space, {\it i.e.} on the
surface $P_4=0$.

The quantities $P_1$ and $M_{\mu\nu}$ make up the Lie algebra with respect to
PB:
\begin{eqnarray} \label{Lee-alg.}
&&\{M_{\mu\nu},P_{1\sigma}\}=-g_{\mu\sigma}P_{1\nu}+g_{\nu\sigma}P_{1\mu},\nonumber\\
&&\{M_{\mu\nu},M_{\rho\sigma}\}=g_{\mu\sigma}M_{\nu\rho}+
g_{\nu\rho}M_{\mu\sigma}-g_{\nu\sigma}M_{\mu\rho}-g_{\mu\rho}M_{\nu\sigma}.
\end{eqnarray}
One can construct the Casimir operators $P_1^2=M^2$ and $W$ of the Poincar\'e
group, where $W$ is determined as
\begin{equation} \label{W}
W=-S^2=\frac{1}{2}M_{\mu\nu}M^{\mu\nu}P_1^2-
M_{\mu\sigma}M^{\mu\tau}P_1^\sigma P_{1\tau}
\end{equation}
with the particle spin defined in the case $d=2+1$ by
\begin{equation} \label{spin}
S=\frac{1}{2}|P_1^2|^{-1/2}\varepsilon_{\mu\nu\rho}P_1^\mu
M^{\nu\rho}.
\end{equation}
Let us obtain the dependence of the particle mass $M$ on parameters of the
curvature and torsion. To this end, we solve the equation of the constraint
$\Psi_2^4$ with respect to $P_1^2$:
\begin{equation} \label{sqrM}
P_1^2=M_{\pm}^2=m^2-\left[\frac{\beta k}{2}\pm
\sqrt{\frac{\beta^2 k^2}{4}+\alpha mk}\right]^2,
\end{equation}
where $k=\sqrt{g}/(q_2^2)^{3/2}$ is the curvature. At a given value of
curvature $k$, the particle has the mass either $M_{+}$ or $M_{-}$. Since
we can impose up to three gauge conditions
\begin{equation} \label{chi}
\chi_\alpha(q,p,\tau)=0,~~~~~~ \alpha=1,2,3
\end{equation}
with the requirements
\begin{eqnarray}
&&{\rm det}||\{\chi_\alpha,\Phi_1^m\}||\not\approx0,~~~~~~m=1,2,3,
\label{chi:det}\\
&&\frac{\partial \chi_\alpha}{\partial \tau}-\{\chi_\alpha,\Phi_1^3\}+
v_m\{\chi_\alpha,\Phi_1^m\}\approx0    \label{chi:requir.}
\end{eqnarray}
($v_m$ are determined from eqs.\ref{chi:requir.} for constructing generalized
Hamiltonian $H_E$), the appearence of the tachyonic sector ($M^2<0$) can be
prevented. Indeed, choosing
\begin{equation} \label{chi:fix}
\chi_1=k+cq_3^2=0,~~~~~\chi_2=q_2q_3=0,~~~~~~~\chi_2=q_1^0-\tau=0,
\end{equation}
we obtain the determinant \ref{chi:det} to be equal to $-4cq_2^0q_2^2q_3^2$.
Then solving the inequality $M^2>0$, one can limit desired values $c$.

For the relation between the mass of the particle and its spin we obtain the
same two-branched expression as in Ref. \cite{Nest.2}
\begin{equation} \label{R-tr}
S=\pm\alpha\sqrt{\left(\frac{m}{M}\right)^2-\theta}-\beta\frac{m}{M} ~~~~~~~
(\theta=\mbox{sign} ~p_1^2),
\end{equation}
{\it i.e.}, for a fixed mass there are two world curves corresponding to
different spin values. \\

\section*{Acknowledgments}
The authors are grateful to V.V.~Nesterenko, V.N.~Pervushin and V.P.~Pavlov
for useful discussions. The authors thanks the Russian Foundation for
Fundamental Research (Grant N$^{\underline {\circ}}$ 96-01-01223) for support.

\end{document}